\documentclass[12pt]{iopart}
\usepackage{bm}
\usepackage{epsfig}
\eqnobysec

\def\l{{\ell}}
\def\lm{{\l m}}
\def\summ{\sum_{m=-\l}^{\l}}
\def\suml{\sum_{\l=0}^{\infty}}

\def\dt{\Delta T}

\def\alm{a_{\lm}}
\def\ylm{Y_{\lm}}

\def\healpix{H{\sc ealpix }}
\def\glesp{G{\sc lesp }}
\def\wmap{\hbox{\sl WMAP~}}

\def\c{{\rm c}}
\def\d{{\rm d}}
\def\planck{{\sl Planck~}}
\def\xcf{{X_\l^{\c f}}}
\def\xdf{{X_\l^{\d f}}}
\def\xdfi{{X_\l^{\d f^i}}}
\def\dilc{{\rm dilc}}
\def\ilc{{\rm ilc}}
\def\d{{\rm d}}

\def\cl{C_{\l}}

\def\cmb{{\rm cmb}}
\def\flmi{{f_\lm^i}}
\def\flm{{f_\lm}}
\newcommand{\nbi}{{Niels Bohr Institute, Blegdamsvej 17,
DK-2100 Copenhagen, Denmark}}
\newcommand{\nott}{{ School of Physics \& Astronomy, University of
Nottingham, University Park, Nottingham NG7 2RD, United Kingdom}}

\begin{document}


\title[The 1DFR and Large Angular Scale Foreground Contamination in the WMAP3 Data]{The One-dimensional Fourier Representation
and Large Angular Scale Foreground Contamination in the 3-year
Wilkinson Microwave Anisotropy Probe data}

\author{Lung-Yih Chiang$^1$, Peter Coles$^{2,3}$, Pavel D Naselsky$^1$, Poul Olesen$^1$}
\address{$^1$ \nbi}
\address{$^2$ \nott}
\address{$^3$ Niels Bohr Institute, Niels Bohr International
Academy, Blegdamsvej 17, DK-2100 Copenhagen, Denmark}

\ead{\mailto{chiang@nbi.dk}, \mailto{peter.coles@nottingham.ac.uk},
\mailto{naselsky@nbi.dk}, \mailto{polesen@nbi.dk}}

\begin{abstract}
We employ the one-dimensional Fourier representation (1DFR) to
analyze the 3-year \wmap de-biased internal linear combination
(DILC) map and its possible contamination by galactic foregrounds.
The 1DFR is a representation of the spherical harmonic coefficients
for each $\l$--mode using an inverse Fourier transform into
one-dimensional curves. Based on the {\it a priori} assumption that
the CMB signal should be statistically independent of, and
consequently have no significant correlation with, any foregrounds,
we cross-correlate the 1DFR curves of $2 \le \l \le 10$ modes, which
are claimed by the \wmap team to be free of contamination and
suitable for whole sky analysis. We find that 8 out of the 9 modes
are negatively cross correlated with the foreground maps, an event
which has a probability of only $9/512 \simeq 0.0176$ for
uncorrelated signals. Furthermore, the local extrema of the 1DFR
curves between the DILC and those of the foregrounds for $\l=2$ and
6 are correlated with significance level below 0.04. We also discuss the minimum variance optimization method and use the properties of the measured cross-correlation to estimate the possible
level of contamination present in the DILC map.
\end{abstract}


\maketitle

\section{Introduction}
The cosmic microwave background (CMB) radiation contains a wealth of
information about our Universe. Not only does the angular power
spectrum of the CMB temperature fluctuations allow us to determine
cosmological parameters that shape our understanding of the
Universe, but the issue of primordial Gaussianity and statistical
isotropy of the CMB also taps the most fundamental principle in
cosmology, and any robust detection of a  deviation would have
far-reaching and indeed revolutionary implications.

Interpreting the term ``Gaussianity'' in the most relevant way for
cosmology, i.e. as meaning that the fluctuations form a
statistically homogeneous and isotropic Gaussian random field \cite{bbks,be}, the
issue of non-Gaussianity in the CMB data was first raised by 
\cite{cobeng} in the COBE data \cite{cobe}. After the release of the
1-year Wilkinson Microwave Anisotropy Probe (\wmap) data
 \cite{wmapresults,wmapfg,wmapng,wmapcl},
 departures from Gaussianity have been detected with various methods
\cite{wmaptacng,gaztanagz,coleskuiper,park,eriksenmf,santanderng,romanng,hansen,mukherjee,larson,phaserandomwalk,mcewen,edingburgh,mnn}.
Although Gaussianity (as we define it) requires statistical
isotropy, a special focus and results on the breaking of large-scale
statistical isotropy have been reported in \cite{toh,dtzh,eriksenasym,copi,schwarz,evil,bernui,abramolow}.
Among them, the most notable anomaly involves the alignment of the
quadrupole and octupole alignment in the general direction of Virgo.
In the 3-year \wmap data \cite{wmap3ytem,wmap3ycos} the previously
detected departures from Gaussianity and statistical isotropy still
persist \cite{wmap3yrng,cruz,mcewenwmap3,copi3y}. The question is whether they are attributed to foreground
residuals \cite{abramo}, systematic errors \cite{dipolestraylight}, local structures \cite{void} or even new physics \cite{jaffebianchi,gordon}.

However, it is important to eliminate the least exciting
possibilities before getting carried away by the most exotic ones. In
particular, one has to be cautious about foreground residuals as
they are surely present in the derived CMB map. The elimination of
foregrounds and the extraction of CMB signal is based on the concept
of minimum variance optimization with the {\it a priori} assumption that CMB and the foregrounds are statistically independent: linearly combining all available maps and
minimizing the variance of the combined map to reduce the foreground
residuals as much as possible. The 3-year de-biased internal linear
combination (DILC) map is based on such a concept by the \wmap
science team. This approach can certainly help to achieve the
optimal estimate of the angular power spectrum for the CMB. It
cannot, however, guarantee the optimal morphology (spatial
distribution). Since most of the anomalies, with the exception of
the reported low quadrupole power \cite{efstathiouquadrupole,dtzh,efstathiou}, are related to the pattern of
fluctuations rather than simply their amplitude, one needs to check
the derived CMB map very carefully before any scientific conclusion
is reached. In this light, several authors have investigated
possible foreground contaminations \cite{foreground,faraday,template,2n} and the effect of different subtraction methods \cite{dt,park3y}.

The {\it a priori} assumption that the CMB and the foregrounds are statistically independent and thus have no significant cross-correlation is not only the backbone of the minimum variance optimization method, it is also the most basic principle (preceding Gaussianity of the CMB), furnishing the most fundamental statistical test for foreground cleaning. Note that statistical independence involves an ensemble of universes, there will, therefore, inevitably be accidental cross-correlations caused by chance alignments, particularly on large angular scales due to the cosmic variance effect \cite{cosmicvariance}. 

Efforts on the investigation on large-scale foreground contamination by direct
cross-correlation were made in the harmonic domain. Based on the
close connection between the phases of the spherical harmonic modes,
and morphology \cite{morph}, \cite{ndv03,ndv04,autox} examine
foreground residuals by cross-correlation of phases between the
\wmap internal linear combination maps and the derived foreground maps.

In this paper, we use a new representation of the spherical harmonic
modes to analyze the cross correlation between the DILC and
foreground maps. Instead of using the the two-dimensional spheres,
we use an inverse Fourier transform on the spherical harmonic modes
for each $\l$ producing one-dimensional curves for each harmonic
scale. We therefore call such representation one-dimensional Fourier
representation (1DFR). Chiang and Naselsky \cite{autox} first
devised the 1DFR to illustrate connection between phase coupling and
morphology, particularly relating to local extrema higher than
3$\sigma$. Chiang, Naselsky and Coles \cite{wmap3yrng} use the same
representation to demonstrate the anomalies in the distribution of
global extrema of the 1DFR curves of the DILC map for the $\l \le
10$ modes.

This paper is arranged as follows. In Section 2 we summarize the
Gaussian random hypothesis of the CMB. In Section 3 we introduce the
1DFR and its properties and advantages. We then use simple
cross-correlation and extremum correlation to examine foreground
contamination in the DILC map in Section 4. We discuss the minimum variance optimization method and use cross-correlation coefficients to estimate the level of foreground contamination in
Section 5. The conclusions are presented in Section 6.

\section{Gaussian Random hypothesis of the CMB}

The statistical characterization of CMB temperature fluctuations
(against the CMB isotropic temperature $T_0=2.725$ K) on a sphere
can be expressed as a sum over spherical harmonics:
\begin{equation}
T(\theta,\varphi)=\suml \summ \alm \ylm (\theta,\varphi),
\end{equation}
where the $\ylm(\theta,\varphi) $ are spherical harmonic functions,
defined in terms of the Legendre polynomials $P_\lm$ using
\begin{equation}
\ylm(\theta,\varphi)=(-1)^m
\sqrt{\frac{(2\l+1)(\l-m)!}{4\pi(\l+m)!}}P_\lm(\cos\theta)\exp(i m
\varphi),
\end{equation}
and the $\alm$ are complex coefficients which can be expressed with
$\alm=|\alm| \exp(i \Phi_\lm)$ and $\Phi_\lm$ are the phases. We use
the Condon-Shortly definition for the spherical harmonic
decomposition. Isotropic Gaussian random CMB temperature
fluctuations on a sphere, of the type that result from the simplest
versions of the inflation paradigm, possess spherical harmonic
coefficients $\alm$ whose real and imaginary parts are mutually
independent and both Gaussian, or equivalently, the amplitudes
$|\alm|$ are Rayleigh distributed with random phases \cite{bbks,be}.
The statistical properties are then completely specified by the
second-order statistics, the angular power spectrum $\cl$,
\begin{equation}
\langle  a^{}_{\l^{ } m^{ }} a^{*}_{\l^{'} m^{'}} \rangle = \cl \;
\delta_{\l^{ } \l^{'}} \delta_{m^{} m^{'}}.
\end{equation}
Since $T$ is always real, the complex vectors of the $\alm$ on the
Argand plane for $m<0$ are mirror images of $m>0$ with respect to
$x$ axis for even $m$, and with respect to $y$ axis for odd $m$. The
statistics for real signal on a sphere are therefore registered only
in the spherical harmonic coefficients $\alm$ for $m\ge0$.

\section{One dimensional Fourier representation of the spherical harmonic coefficients}
Recent studies of the large angular scale properties for the CMB
temperature anisotropy \cite{toh,dtzh,schwarz,evil} are based on
the composite map constructed for each $\l$ by summing all the $m$
modes pertaining to that $\l$:
\begin{equation}
T_\l(\theta, \varphi) = \sum_{m=-\l}^\l \alm \ylm(\theta,\varphi).
\label{composite}
\end{equation}
Alternatively, one can represent $\alm$ in each $\l$ by an inverse
Fourier transform, as the $\alm$ is now a function of a single
variable $m$. Such a 1DFR for $\alm$ from the spherical harmonic
decomposition of the sky is written as
\begin{equation}
T_\l(\varphi)=\sum_{m=-\l}^\l \alm \exp(i m \varphi)=a_{\l 0}+2\sum_{m=1}^\l |\alm| \cos(m\varphi+\Phi_\lm),
\end{equation}
where for the negative $m$ we use the complex conjugate
($a_{\l,-m}=\alm^\ast$) to make $T_\l(\varphi)$ real. As we have
mentioned for real $T$ on a sphere the statistics are registered
only in $m \ge 0$ modes, the $T_\l(\varphi)$ curves assembled in
this way contain the same amount of information as the spherical
$T_\l(\theta, \varphi)$.  The variance for each curve is then
\begin{equation}
\sigma_\l^2=\frac{1}{2\pi}\int_0^{2\pi} \left[T_\l(\varphi)-\overline{T_\l}\right]^2 d\varphi=2\sum_{m=1}^\l |\alm|^2.
\end{equation}
For a Gaussian Random Field (GRF), the $|\alm|$ have a Rayleigh
distribution, so $\sigma_\l^2$ has a chi-square distribution.

\subsection{Comparison between 1DFR and integration on $\theta$ of the composite maps}
\begin{figure}
\hspace*{25mm}
\epsfxsize=10cm
\epsfbox{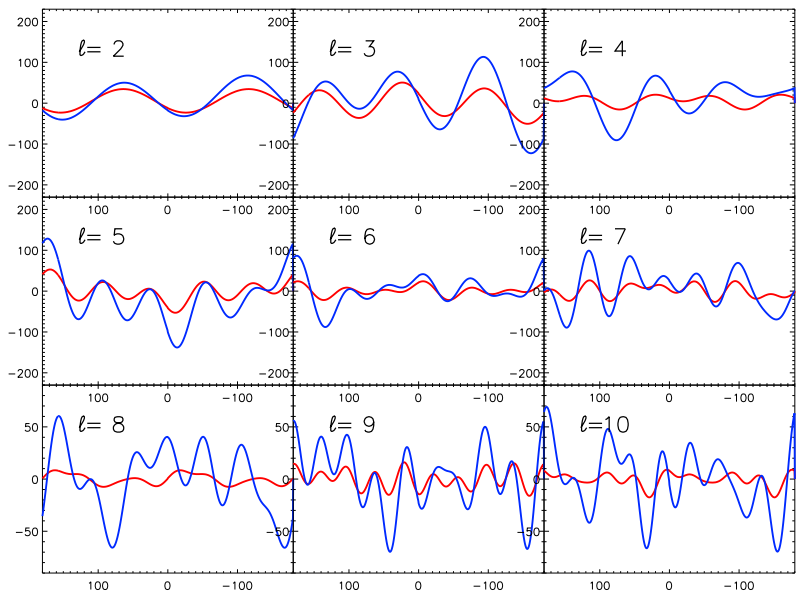}
\caption{Comparison of the 1DFR of the DILC map (blue curves) and the $T$ distribution curves from integration of $T_\l(\theta,\varphi)$ on $\theta$ direction (red curves). The $x$ axis is $\varphi$ and is plotted reversely to follow the conventional Galactic longitude coordinate $l$ and the $y$ axis is in unit of thermodynamic temperature ($\mu$K). Note that the amplitude of the 1DFR curves for odd $\l$ is multiplied by $-1$ in order for comparison with those from integration on $\theta$.}\label{fig:compare}
\end{figure}

Since the 1DFR results in $\dt$ as a single function of $\varphi$,
such representation is therefore similar to the integration on
$\theta$ used in the composite maps:
\begin{equation}
\int_0^\pi T_\l(\theta,\varphi) \; d\theta.
\end{equation}
There are, however, some subtleties. \Eref{composite} can be written
as follows
\begin{eqnarray}
 T_\l(\theta,\varphi) = \sqrt{\frac{2\l+1}{4\pi}} \left\{ \frac{}{} a_{\l 0} P_{\l 0}(\cos\theta) \right.\nonumber \\
+ \left. 2\sum_{m=1}^\l  \left[ (-1)^m  \sqrt{\frac{(\l-m)!}{(\l+m)!}} \left|\alm \right|\cos(m\varphi +\Phi_\lm) P_\lm(\cos \theta)\right]\right\}.
\end{eqnarray}
As $\int P_\lm(\cos \theta) d\theta =0 $ for odd $\l+m$, the
curves from the integration over $\theta$ from the composite maps
have contributions from only the odd $m$ part for odd $\l$ and the
even $m$ part for even $\l$. Therefore for odd $\l$, the $(-1)^m$
term inverts the resulting temperature $T$ as seen from the 1DFR.
(As we use the Condon-Shortley definition for the spherical harmonic
functions, $\int P_\lm(\cos \theta)d\theta$ is always positive for
$m>0$ and 0 for $m=0$). Accordingly, in \Fref{fig:compare}, for the
odd $\l$ curves the amplitudes are multiplied by $-1$ to facilitate
direct comparison. The $\varphi$ coordinate of the 1DFR is also
plotted backwards in order to match the Galactic longitude
coordinate $l$ (not to be confused with the spherical harmonic mode
$\l$; see \Fref{fig:map} (top panel).

Moreover, the integration over $\theta$ results in more symmetric
curves than the simple 1DFR, where equal footing is given to each
$\alm$ in the summation. As shown in \Fref{fig:compare}, the curves
(red) obtained by integration of $\theta$ from the composite maps
show repetitions for $\varphi $ at $[0, 180^\circ]$ and
$[-180^\circ, 0]$.

\subsection{Characteristics of the 1DFR}
The 1DFR assembles spherical harmonic coefficients into
one-dimensional curves, so the information on $\theta$ direction is
lost, compared with standard two-dimensional spherical maps.
Nevertheless, the morphology and statistics registered in the
complex $\alm$ sequence should still manifest themselves in this
representation. For a Gaussian random field on a sphere, the real
and imaginary parts of the spherical harmonic coefficients $\alm$ in
each $\l$ are mutually independent and both Gaussian distributed
with zero mean and variance $\cl/2$, where $\cl$ is its angular
power spectrum. Thus if the $\alm$ are a result of Gaussian process,
the 1DFR $T_\l(\varphi)$ curves shall possess all the usual Gaussian
random properties such as two-point correlation, peak statistics,
and Minkowski functionals\ldots etc..

\begin{figure}
\hspace*{25mm}
\epsfxsize=10cm
\epsfbox{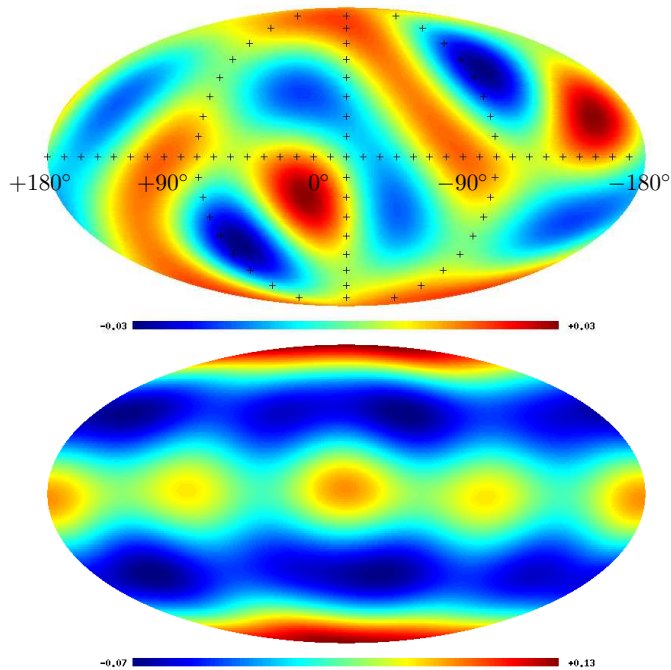}
\caption{The composite map of $\l=4$ from DILC map (top) and that of $\l=4$ from the W channel foreground map (bottom). One can see the dominant stripes of the foreground composite map caused by $a_{\l 0}P_{\l 0} (\cos \theta)$, which makes it difficult to be compared with the DILC composite map.}
\label{fig:map}
\end{figure}

One of the 1DFR characteristics is that the $a_{\l 0}$ modes
contribute to the 1D curves only a constant shift without altering
the morphology, whereas in spherical harmonic composition it
produces modulation in $\theta$ direction, $a_{\l 0} P_{\l 0}(\cos
\theta)$. This is useful when one is to compare two sets of $\alm$
with standard cross-correlation, particularly when one of them is
from a foreground map. Standard analysis on large angular scale (low
multipole) anomalies is performed on a composite map by
\Eref{composite}: a full-sky map synthesized from the $\alm$ from
$-m$ to $m$ for each $\l$ (see \Fref{fig:map} top panel as an
example). For foreground maps defined in Galactic coordinate system,
the emission near the Galactic plane dominates the signal, which is
then spherical-harmonic transformed into high amplitudes of
$a_{\l0}$ for even $\l$ (\Fref{fig:map} bottom), which produce a
prominent belt in the composite maps, making them rather difficult
to analyze under such situation.

\section{1DFR and Foreground contamination in the DILC map}
In this section we investigate foreground contamination in the \wmap
3-year DILC map. It is based on the {\it a priori} assumption that
the CMB (at the background) should have no ``knowledge'' about what
the foregrounds look like. So if the derived DILC map more or less
reflects the morphology of true CMB, the DILC and foregrounds should
have little or no resemblance to it; we can also see how clean is
the derived DILC by directly comparing the morphology.

\begin{figure}
\hspace*{25mm}
\epsfxsize=10cm
\epsfbox{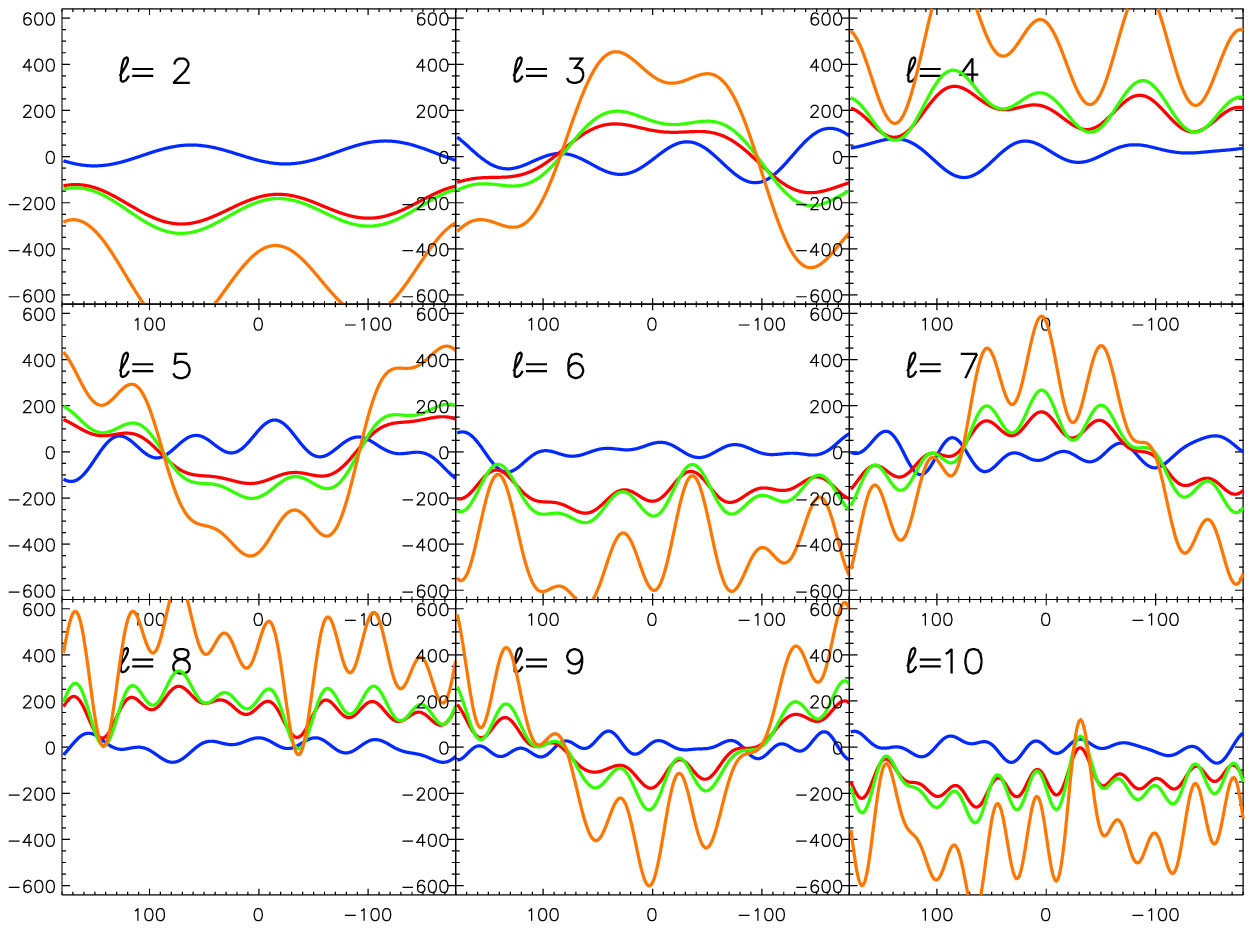}\\
\hspace*{25mm}
\epsfxsize=10cm
\epsfbox{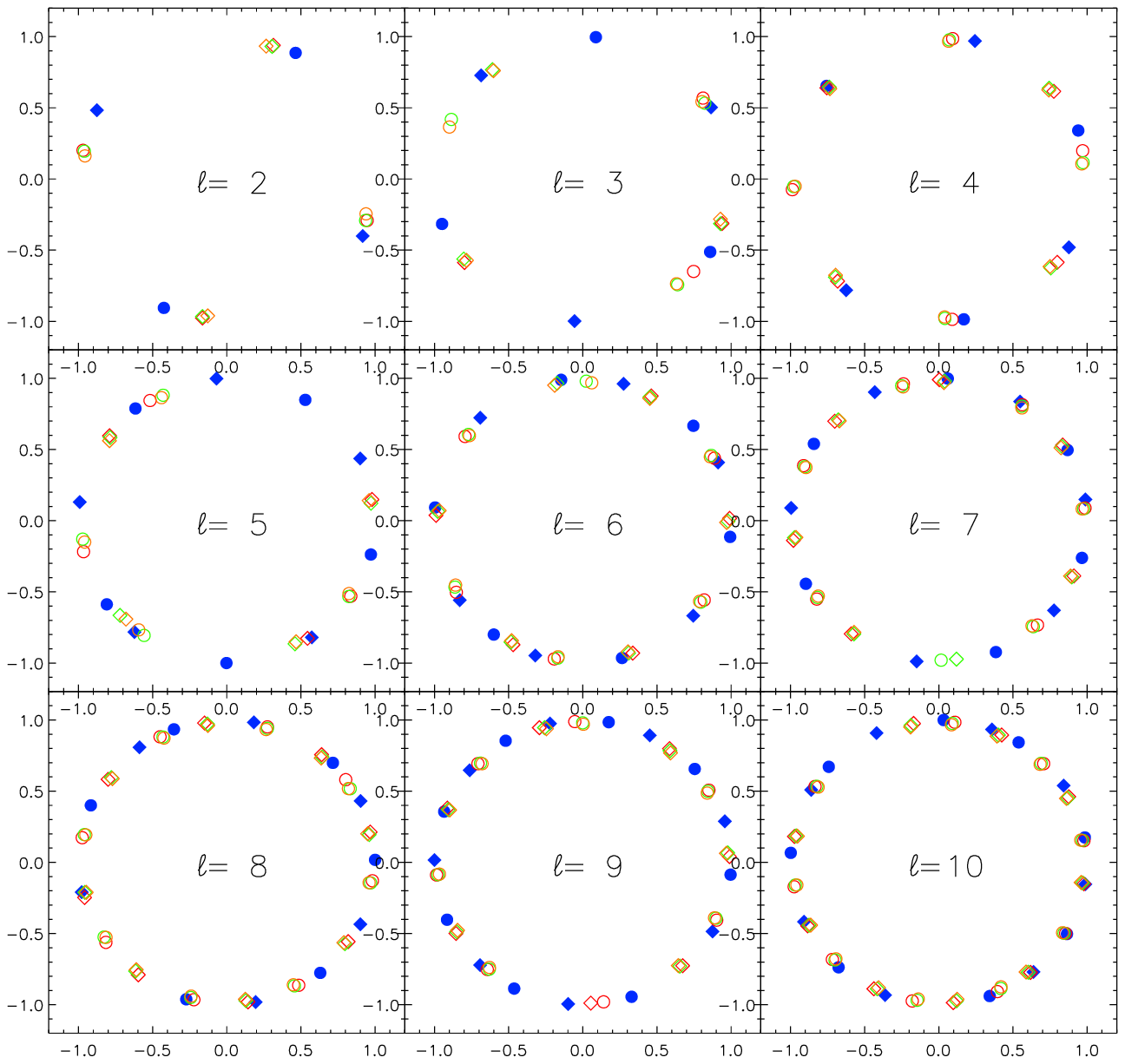}
\caption{Top mosaic: the 1DFR curves for DILC (blue curves) and those of the foreground (sum of the synchrotron, free-free and dust templates) maps at Q channel (orange curves), V channel (green curves) and W channel (red curves). The $x$ axis is $\varphi$ and is plotted reversely to follow the conventional Galactic longitude coordinate $l$ and the $y$ axis is in unit of thermodynamic temperature ($\mu$K). Notice strong anti cross correlation for the quadrupole. Bottom mosaic: the position of the local extrema plotted on unit circles. The angle between each point and the positive $x$ axis is the position $\varphi$ for each extrema of the 1DFR curves from the DILC map (blue filled sign) and those from foreground map at Q channel (orange open sign), V (green open sign) and W channel (red open sign). The circle and diamond signs represent peaks (local maxima) and troughs (local minima), respectively. The signs subtend about $5^\circ$ so that, for example, one can see for $\l=10$ there are 12 out of totally 16 local extrema of
DILC 1DFR curves located within $5^\circ$ of those of W channel foreground curve.} \label{1dfrall}
\end{figure}

Firstly, in the top mosaic of \Fref{1dfrall} we plot the 1DFR curves
for the DILC (blue curves) and those of the \wmap derived foreground
maps at Q (orange), V (green) and W  (red) channels. The foreground
maps are the sum of the synchrotron, free-free and dust templates
obtained via Maximum Entropy Method \cite{wmap3ytem}. As one can
see, the most striking feature is the significant anti-correlation
between the DILC and foregrounds for the quadrupole ($\l=2$).

In the bottom mosaic of \Fref{1dfrall} we plot, on a unit circle for
each 1DFR curve, the positions of local extrema: all $\varphi$ for
$dT_\l/d\varphi=0$ and $d^2T_\l/d\varphi^2\ne0$. The angle between
each sign and the positive $x$ axis is $\varphi$ for each local
extremum position. The blue filled signs are from the DILC and the
open signs the foreground map at Q (orange), V (green) and W (red)
channels. We use circles and diamonds to denote peaks
($d^2T_\l/d\varphi^2 <0$) and troughs ($d^2T_\l/d\varphi^2 >0$),
respectively. The size of the signs is $\simeq 5^\circ$ so that one
can see there are 12 out of totally 16 local extrema for $\l=10$
DILC 1DFR curves with W channel foreground map 1DFR local extrema
located within $5^\circ$.

In what follows, we employ cross correlation and correlation of
extrema of the 1DFR curves to illustrate the foreground
contamination in the DILC map. The cross correlation is a measure of
``trend'': two curves will yield strong cross correlation if both go
up or down in tandem at the same interval of $\varphi$. However, any
correlation will be cancelled out if they match on one half of
$\varphi$, but on the other they are opposite. Although cross and
extremum correlation are not totally independent, it is therefore
intuitively helpful to examine both properties.

\begin{figure}
\hspace*{25mm}
\epsfxsize=13cm
\epsfbox{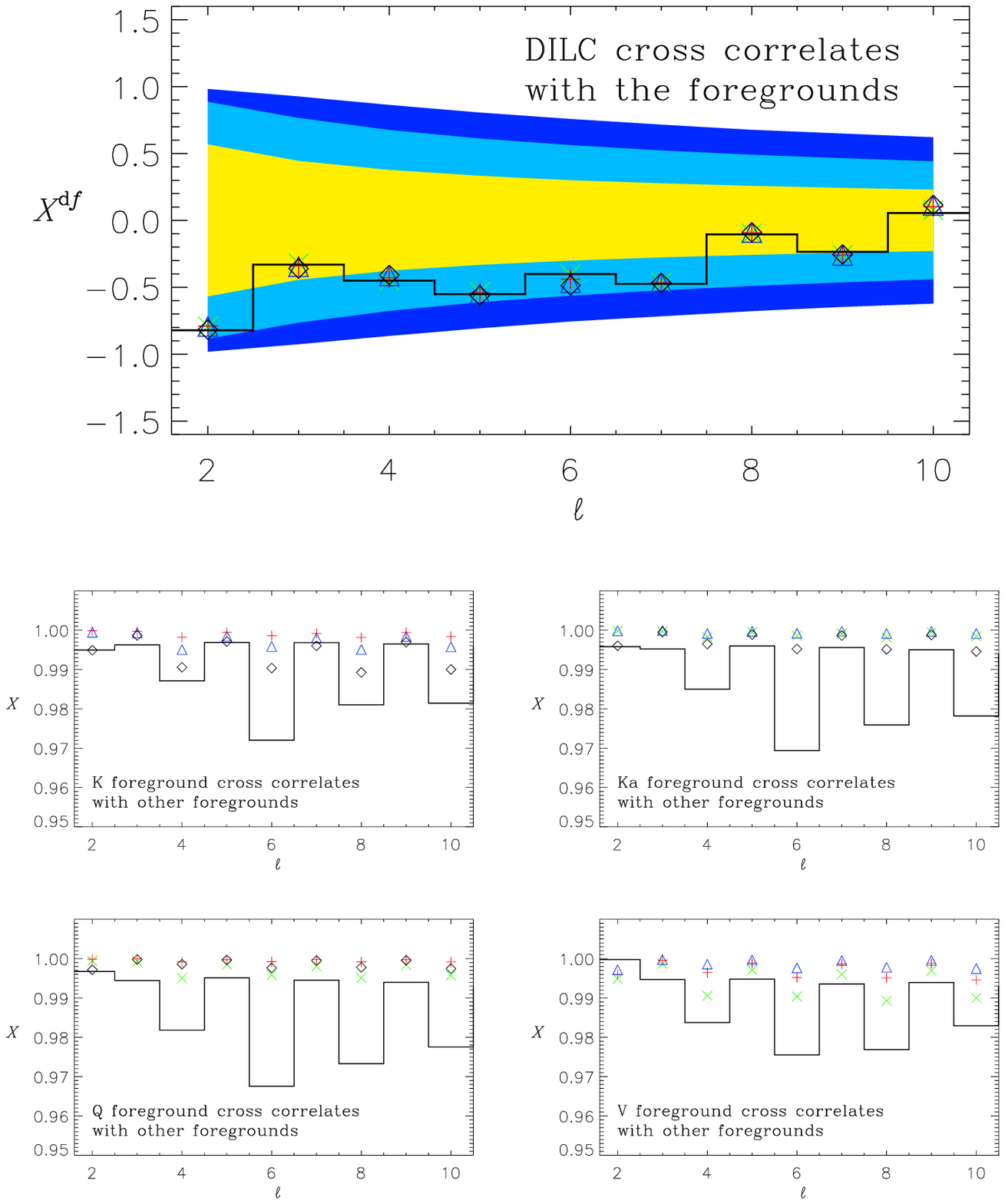}
\caption{Cross correlation of the 1DFR curves from the DILC map with those from \wmap foreground maps (top panel) at K channel (green $\times$), Ka channel (red $+$ signs), Q (blue $\opentriangle$ signs), V (black $\opendiamond$), W (black line) and cross correlation of the 1DFR curves between the foregrounds (bottom 4 panels). Note that 8 out of 9 modes are negatively correlated, the probability for which to happen is $C^9_1 (2)^{-9}\simeq 0.0176$. The yellow, light blue and dark blue areas denote 1, 2 and 3 $\sigma$ respectively (68.27\%, 95.45\% and 99.73\%) of $10^5$ Monte Carlo simulations.} \label{x}
\end{figure}

\subsection{Cross correlations of the DILC map with the \wmap derived foreground maps}
We use cross-correlation to quantify the foreground contamination in
the DILC map:
\begin{equation}
X^{ij}_\l=\frac{ (2\pi)^{-1}\int_0^{2\pi}\left [ T_\l^i(\varphi)-
\overline{T_\l^i} \right] \left[ T_\l^j(\varphi)-
\overline{T_\l^j}\right]d\varphi}{\sigma^i_\l \sigma^j_\l},
\label{xcorr}
\end{equation}
where $T^i_\l(\varphi)$ indicates the 1DFR curve of multipole number
$\l$ at $\varphi$ of $i$ map, $\sigma^i_\l$ is the standard
deviation of the curve. The $X$ coefficients range from $-1$ to 1.
In \Fref{x} we show the $X$ coefficients between 1DFR curves of the
DILC and those of the foregrounds. First of all, if the DILC map is
uncorrelated with the foregrounds, the probabilities for positive
and negative $X$ should be equally $1/2$. One can see that eight out
of the nine modes (except $\l=10$) $X<0$, an outcome for which the
probability is $C^9_1 (2)^{-9}\simeq 0.0176$.

\begin{figure}
\hspace*{25mm}
\epsfxsize=13cm
\epsfbox{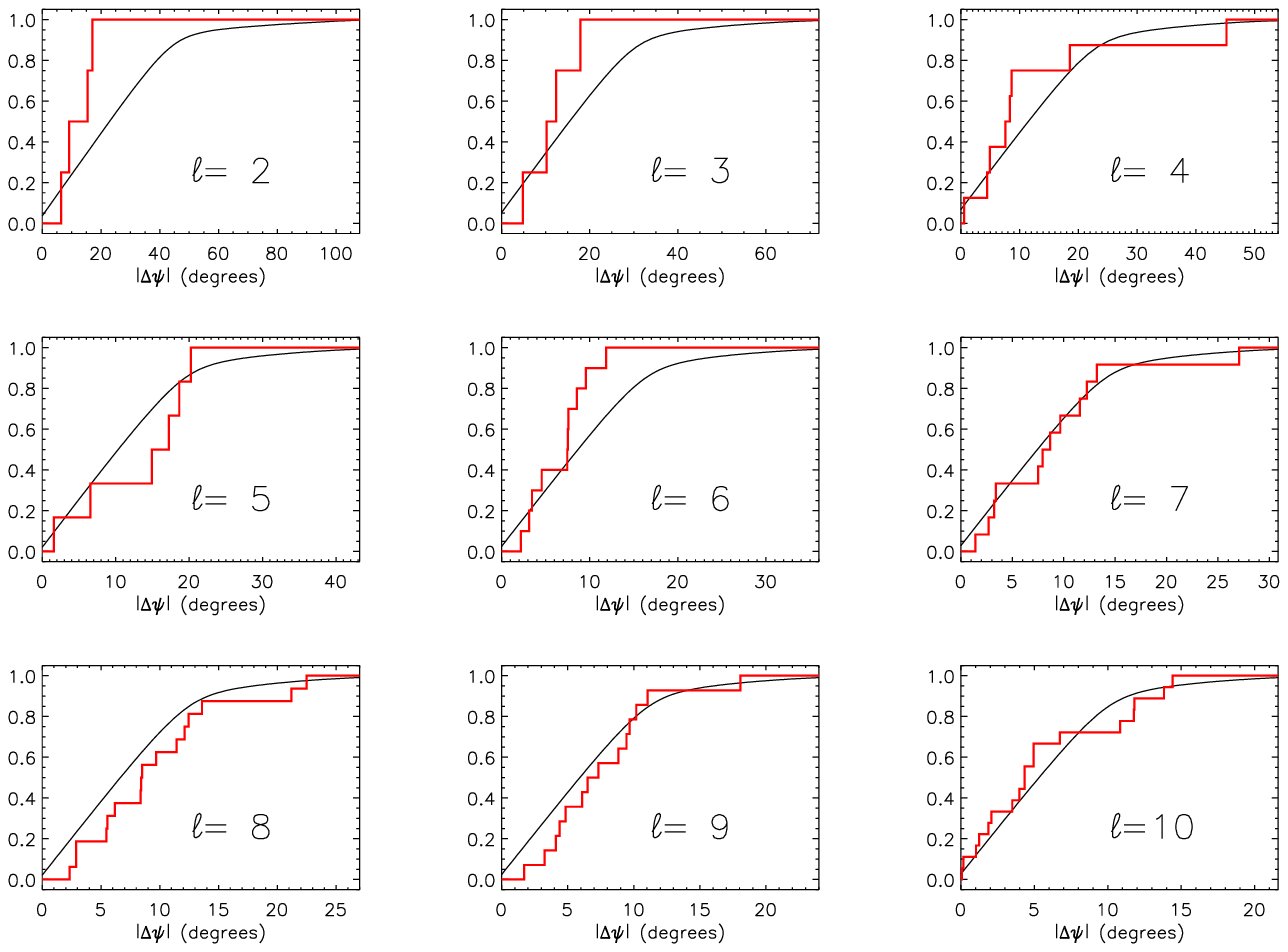}\\
\hspace*{25mm}
\epsfxsize=13cm
\epsfbox{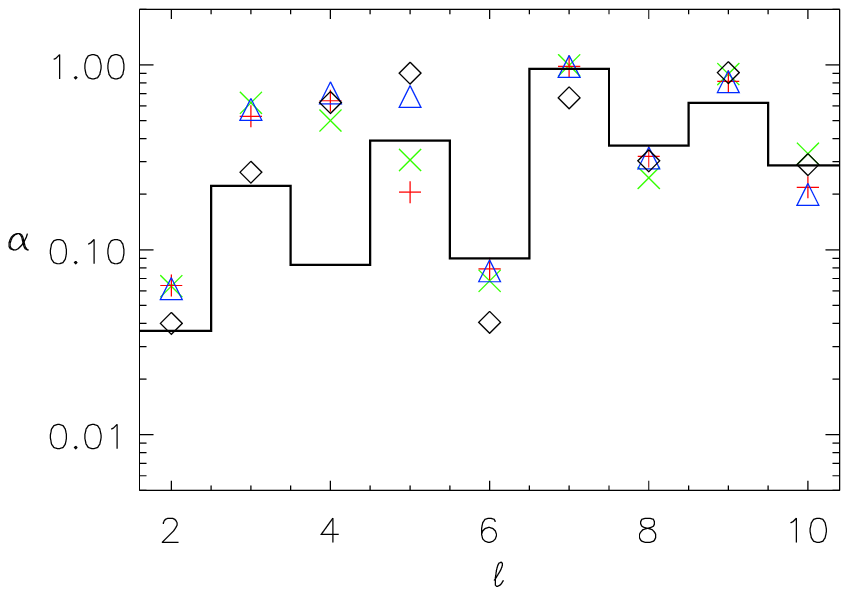}
\caption{Top mosaic: the cumulative probability distribution of the separation angles
(the angle of the nearest extremum) between the DILC and the W channel
foreground extrema is plotted in red curves and the cumulative probability distribution of the separation angles from $10^5$ Monte Carlo simulation is in black. The bottom panel is the significance level from K-S test for the distribution of extrema of DILC with foreground maps at K channel (green $\times$ signs), Ka channel (red $+$ signs), Q (blue $\opentriangle$ signs), V (black $\opendiamond$ signs), W (black line). For $\l=2$ and 6, the significance levels are below 0.04 for the V channel, which is also the most weighted channel by \wmap in internal linear combination of multi-frequency cleaning to extract the CMB signal.} \label{peakx}
\end{figure}

Secondly, the cross correlation in quadrupole is below $-0.8$ (at around 90\% CL). In our previous publication \cite{wmap3yrng}, we use cross correlation of phases between the DILC and foreground maps to examine foreground contamination. The most prominent correlation appears in octupole, which renders the significance level as low as $0.06$. Although we do not claim that the quadrupole-octupole alignment is directly caused by the foregrounds, we can at least claim that since there is significant foreground contamination, such alignment is not cosmological. In the lower 4 panels of \Fref{x} we also show the cross correlation between the foregrounds, which are all above 0.96.

\subsection{Correlation of extrema of the 1DFR curves}
In our previous publication \cite{wmap3yrng}, we used a 1DFR on the
9 modes $2\le \l \le 10$ of the DILC map and examined the positions
of the global extrema of the 9 curves. If the signal were Gaussian,
the global extrema should distribute randomly on the $\varphi$ axis.
We found that the extrema are concentrated around
$\varphi=180^\circ$ and avoid the $\varphi=0$ (Galactic centre) with
significance level below 0.5\%, which is a strong indication that
the global extrema of the DILC map are non-randomly distributed due
to some influence related to a Galactic-coordinate frame. Here we
further examine the local extrema of the 1DFR curves of the DILC map
and their correlation with those of the foregrounds.

In order to extract information about correlation of extrema between
DILC and the foregrounds, we use the following method: for each
foreground extremum, we search for its nearest DILC extremum and
denote their distance by $|\Delta \varphi_\l(i)|$, where $i$ denotes
the $i$th extremum of the foreground curve. The collection of the
separation angles for each $\l$ is to be compared with the
distribution of $10^5$ Monte Carlo simulation on the GRF extrema
against the foreground ones. In the top mosaic of \Fref{peakx} we
show the cumulative probability distribution of the separation
angles between the DILC and W channel foreground extrema (red
curves) and that of separation angles of $10^5$ realizations of
simulated GRF and W channel foreground extrema (black).  The bottom
panel is the significance level from Kolmogorov-Smirnov (K-S) test
for the distribution of extrema of DILC with foreground maps at K
channel (green $\times$ signs), Ka channel (red $+$ signs), Q (blue
$\opentriangle$ signs), V (black $\opendiamond$ signs), W (black
line). One can see that for $\l=2$ and 6, the significance levels
are all below 0.1 for all channels, and specifically, the
significance levels are below 0.03 for the V channel, which is also
the  channel with the highest weight assigned by \wmap in the
internal linear combination of multi-frequency cleaning to extract
the CMB signal.

\subsection{Cross correlation with foreground templates independent from \wmap
data }
\begin{figure}
\hspace*{25mm}
\epsfxsize=13cm
\epsfbox{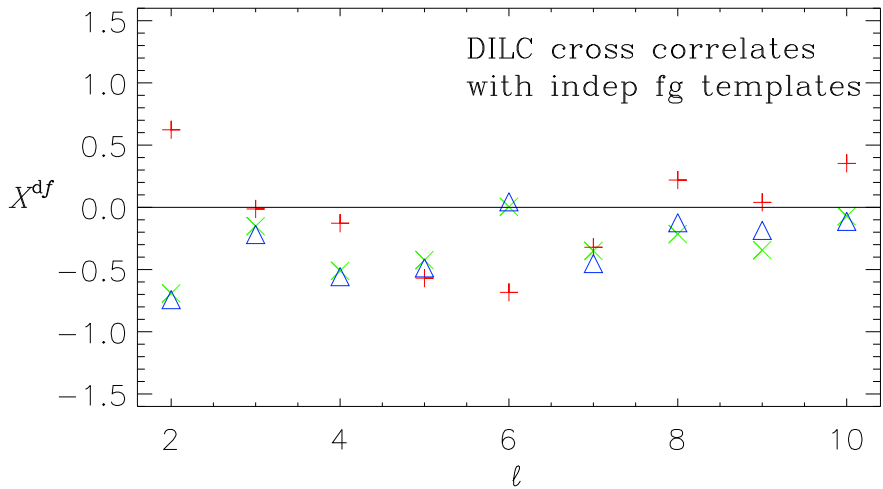}
\caption{Cross correlation of the 1DFR curves from the DILC map with those
  from full-sky foreground templates that are observed, collected or derived
  independently from the \wmap data. These independent templates are the 408
  MHz Continuum Survey (green $\times$) for synchrotron emission, H$\alpha$
  map for free-free emission (red $+$ signs), and the dust template extrapolated from COBE/DIRBE data (blue $\opentriangle$ signs). For the syncrotron and dust
  templates, 8 of the 9 modes are negatively correlated with the DILC, the
  probability is $C^9_1 (2)^{-9}\simeq 0.0176$.} \label{x.indep}
\end{figure}

To further certify foreground contamination in the DILC map, we also test 
various other full-sky foreground templates that are measured, collected or derived 
independently from the \wmap data. The existing foreground templates
are the 408 MHz radio continuum survey \cite{haslam} tracing the synchrotron
pattern, the H$\alpha$ template for thermal bremsstrahlung (free-free
emission) \cite{finkbeiner}, and the dust template extrapolated from
COBE/DIRBE data to microwave frequencies \cite{fds1999}. We repeat the cross
correlation of the 1DFR curves in Section 4.1 and show the cross correlation
coefficient $X$ in Fig.\ref{x.indep}. For both the synchrotron and dust templates,
8 of the 9 modes are negatively correlated with the DILC, the probability
of this happening in the absence of real correlation only being $C^9_1
(2)^{-9}\simeq 0.0176$. We find no significant evidence of cross-correlation
with the H$\alpha$ template, but this is not unexpected since this makes only
a small contribution to the expected foreground at relevant frequencies.
One should note that, although the 408 MHz
map and the dust template are templates relating to single foreground components, they
can be considered as total foreground templates as they represent the dominant
component at the corresponding frequencies. Their correlation with the
DILC strengthens the claims we made in the previous section. 

\section{Cross correlations with the foregrounds as an estimate on the foreground contamination}
\subsection{Minimum variance optimization and foreground residuals}
In the last section, we have shown through strong cross correlation
and peak correlation that the DILC signal is probably contaminated
with foregrounds. We now examine the concept of multi-frequency
cleaning method and how foreground residuals propagate into the
final map. Multi-frequency cleaning has been the workhorse for
retrieving the CMB signal from polluted data. It is based on the
concept that the CMB signal exists, among different frequency maps,
as a frequency independent component \cite{wmapfg,wmap3ytem} (see also \cite{lilc}):
\begin{equation}
T^i=T^\cmb + F^i,
\label{freqmap}
\end{equation}
where $T^i$ represents the total measured signal at frequency band
$i$, $T^\cmb$ the frequency independent (i.e. the CMB) signal, and
$F^i$ the foreground at frequency band $i$ (here we assume noise is
not important). This frequency-independent component can be flushed
out with an internal linear combination on $M$ frequency bands of
maps with weighting coefficients $\sum_{i=1}^M w_i=1$:
\begin{equation}
T^\ilc=\sum_{i=1}^M w_i T^i = T^\cmb + \sum_{i=1}^M w_i F^i,
\label{minivar}
\end{equation}
where $T^\ilc$ represents the ILC map. The {\it a priori} assumption that the
frequency independent component should be statistically independent
with others ensures the variance of the ILC map is the sum of the
variances of the two parts:
\begin{equation}
{\rm var}[T^\ilc]={\rm var}[T^\cmb]+{\rm var}[\sum_{i=1}^M w_i F^i].
\label{ilc}
\end{equation}
Therefore if one is to minimize the variance of the ILC map by $\partial {\rm var}[T^\ilc]/\partial w_i =0$, it is
equivalent to minimizing the variance of linear combination of the
foregrounds, 
\begin{equation}
\frac{\partial {\rm var}[T^\ilc]}{\partial w_i} \equiv \frac{\partial {\rm var}[\sum_{i=1}^M w_i F^i]}{\partial w_i},
\end{equation}
thus reducing most the residuals to produce a map closest to
the CMB. The \wmap DILC map is thus produced by employing the
minimum variance optimization in the pixel domain in 12 separate regions, where Region
0 marks the largest region for $|b| \geq 15^\circ$, and Region
$1-11$ for those around the Galactic plane. In each region the 5
frequency maps are linearly combined and a set of weighting
coefficients $w_i^{(R)}$ are obtained in such a way that the
resultant variance is minimum.

It is worth noting that \Eref{ilc} requires statistical independence that is defined between the foregrounds and an ensemble of universes, but the fact that we have only one universe shall introduce some error in the minimization process. 
Furthermore, since the minimum variance optimization takes on the {\it overall} variance of the map,
instead of the individual multipoles as the method by
\cite{te96,toh,dt}, there is no guarantee that the contamination
from the foregrounds in each multipole $\l$ will be minimum (let
alone zero). A spherical harmonic decomposition of \Eref{minivar}
gives
\begin{equation}
\alm^\dilc=\alm^\cmb+\sum_i w_i \flmi,
\end{equation}
where $\flmi$ represents the spherical harmonic coefficients for the
foreground map $F^i$. For higher multipoles, the power spectrum
$\cl$ of the DILC should be close to that of CMB as there are more
$m$ modes participating in the summation $(2\l+1)^{-1}\sum_m
|\alm^\dilc|^2$, as long as the foreground residual $\sum w_i
f_\lm^i$ is more or less non-correlated between each $m$ of the same
$\l$ (like white noise). On the other hand, for lower multipoles,
the discrepancy between CMB and the DILC can be prominent when the
foreground residuals are correlated.

From the significant cross and extremum correlation as shown in
\Fref{x} and \Fref{peakx}, it is indeed likely that the foreground
residual is still a non-negligible part for low multipoles in the
DILC map, unless the true CMB signal happens to 'resemble' the
foregrounds, an eventuality which has a small probability.

\subsection{Estimate of the foreground contamination}
To make an estimate of the foreground contamination present in the
DILC map, we can first of all write
\begin{equation}
\flmi=\sigma_\l^{f^i} \flm,
\end{equation}
where $\sigma_\l^{f^i}$ is the r.m.s. of the fluctuation of $\flmi$ and
$\flm$ is the {\it unitary} foreground signal (related to
morphology) with its r.m.s. of fluctuation $\sigma_\l^f=1$. Such a
degeneracy can be seen in \Fref{x} that most cross-correlations
between foregrounds are above 0.99 (except with the W channel $\sim
0.97 $).

We now calculate the cross correlation coefficient of the 1DFR
curves between the true CMB and the unitary foreground signal:
\begin{eqnarray}
\xcf&=&\frac{1}{2\pi} \int_0^{2\pi} \frac{ \sum_{m\ne 0} \alm^\cmb \exp(i m \varphi)  \sum_{m^{'}\ne 0} f_{\l m^{'}} \exp(i m^{'} \varphi) }{\sigma_\l^\cmb \sigma_\l^{f}}  d\varphi \nonumber \\
&=&\frac{2\sum_{m=1}^\l  |\alm^\cmb||\flm|\cos(\Phi^\cmb_\lm-\Phi^f_\lm)}{\sigma_\l^\cmb}.
\label{xcmbf}
\end{eqnarray}
Although the CMB and the foregrounds should be statistically independent, i.e. $\langle \xcf \rangle=0$,
the fact that we only have one realization of the CMB fluctuation makes this parameter crucial,
especially for low multipoles.

\begin{figure}
\hspace*{25mm}
\epsfxsize=10cm
\epsfbox{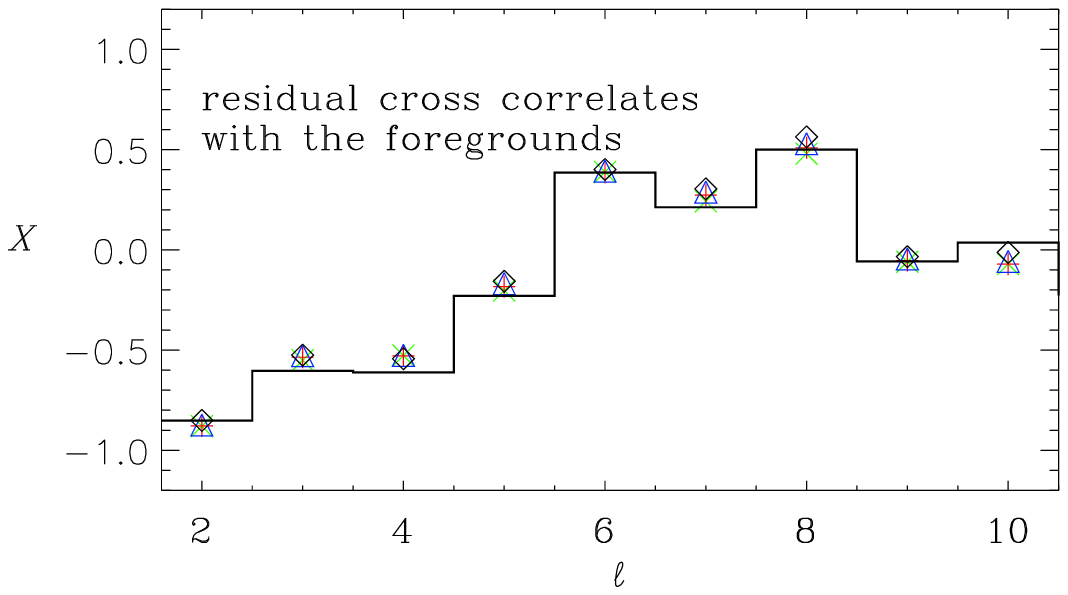} \\
\hspace*{25mm}
\epsfxsize=10cm
\epsfbox{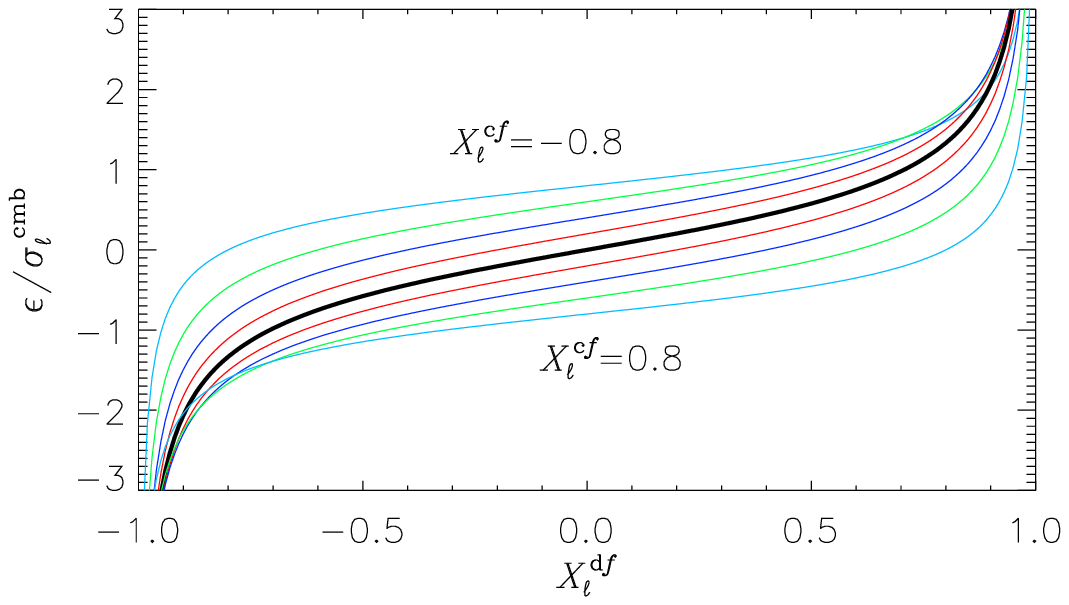} \\
\hspace*{25mm} \epsfxsize=10cm \epsfbox{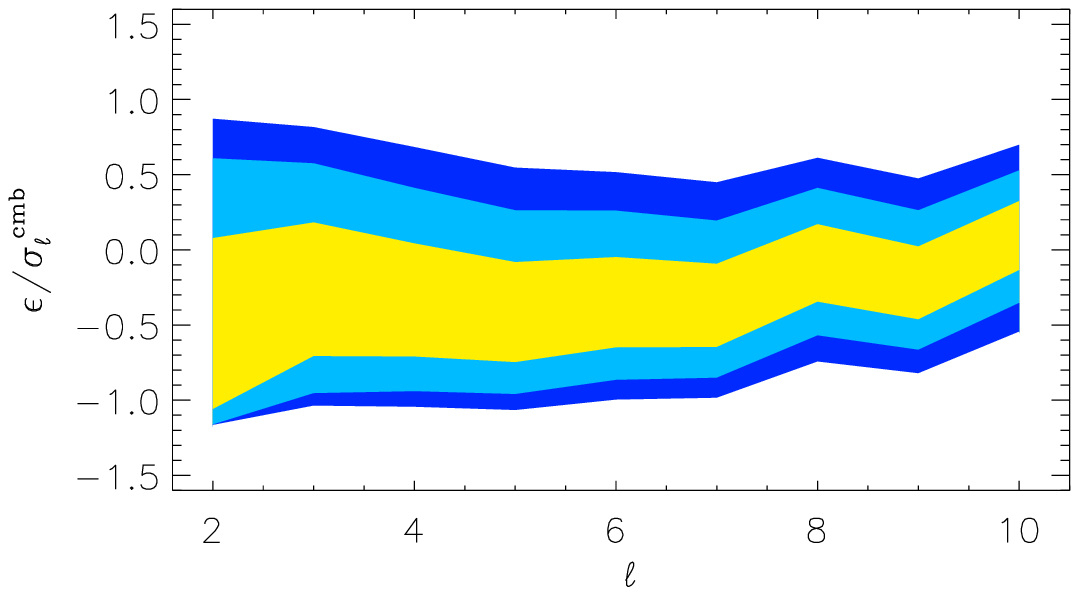} \caption{The
top panel shows the cross-correlation between the foreground
residuals $R_\l$ and the foregrounds $F^i_\l$ and the other two are
the foreground contamination parameter $\epsilon_\l$ in units of
$\sigma_\l^\cmb$. Middle panel is the contamination ratio $\epsilon_\l/\sigma_\l^\cmb$ against the cross correlation coefficients $\xdf$ for DILC and the foregrounds (from \Eref{relation}). The curves are for
$\xcf=-0.8$ (the top) downwards with increment 0.2 to $\xcf=0.8$.
The thick line is for zero correlation: $\xcf=0$ between CMB and the
foregrounds. In the bottom panel, we take the mean value of $\xdfi$ from top panel of \Fref{x} as input and perform $10^6$ Monte
Carlo simulations for the CMB and the $\xcf$. The yellow, light blue and dark blue areas illustrate the
contamination ratio $\epsilon_\l/\sigma_\l^\cmb$ at 1, 2 and 3 $\sigma$ thresholds, respectively.} \label{epsi}
\end{figure}

To relate the $\xcf$ to the measured $X_\l^{\d {f^i}}$, we first
show, in the top panel of \Fref{epsi}, the cross-correlation of the
foreground residuals with the foregrounds. The DILC map is processed
with 12 sets of weighting coefficients $w_i^{(R)}$, where $i=1-5$
represent K, Ka, Q, V and W bands, respectively and $(R)$ indicates
Regions $0-11$, so for the foreground residual we use a linear
combination of the \wmap foregrounds on 12 separated regions to
obtain the whole sky map:
\begin{equation}
R(\theta,\varphi)=\sum_{R=0}^{11} \sum_{i=1}^5 w_i^{(R)} F^i_{(R)},
\end{equation}
from which we produce the 1DFR curves $R_\l$. One can see that for
$\l=2$, 3 and 4, the foreground residual has significant
anti-correlation with the foregrounds. We caution that the \wmap
foregrounds are retrieved via the Maximum Entropy Method, not by
direct subtraction of the DILC signal off the frequency maps, so the
DILC is not a simple sum of the true CMB and the foreground residual
as \Eref{ilc}. For the modes in which the residual has significant
correlation, we express the residuals in terms of the unitary
foreground signals $\epsilon_\l \flm$:
\begin{equation}
\alm^\dilc\simeq\alm^\cmb+\epsilon_\l \flm,
\end{equation}
where $\epsilon_\l$ is the contamination parameter at multipole number $\l$. For other
multipoles, this approach means we are estimating the most possible
contamination.

The cross correlation between the 1DFR curves of the DILC and the
foreground maps can be written as
\begin{eqnarray}
X_\l^{\d f^i}&=&\frac{1}{2\pi} \int_0^{2\pi} \frac{ \sum_{m\ne 0} \alm^\dilc \exp(i m \varphi)  \sum_{m^{'}\ne 0} f^i_{\l m^{'}} \exp(i m^{'} \varphi) }{\sigma_\l^\dilc \sigma_\l^{f^i}}  d\varphi \nonumber \\
&=&\frac{\epsilon_\l+ 2\sum_{m=1}^\l  |\alm^\cmb||\flm|\cos(\Phi^\cmb_\lm-\Phi^f_\lm) }{\sqrt{\epsilon_\l^2+(\sigma_\l^\cmb)^2+4 \epsilon_\l \sum_{m=1}^\l|\alm^\cmb||\flm|\cos(\Phi^\cmb_\lm-\Phi^f_\lm)}}.
\label{relation0}
\end{eqnarray}
We can now denote $X_\l^{\d {f^i}}\equiv \xdf$
as it is independent of the individual foreground channels.
The measured $X_\l^{\d {f^i}} $, as shown in the top panel of \Fref{x},
are indeed degenerate due to the resemblance in morphology between the foregrounds.
With \Eref{xcmbf} we can solve \Eref{relation0} for $\epsilon_\l$:
\begin{equation}
\epsilon_\l= \sigma_\l^\cmb \left(-\xcf + \xdf \sqrt{\frac{1-(\xcf)^2}{1-(\xdf)^2}}\right).
\label{relation}
\end{equation}
\Eref{relation} describes the contamination parameter in relation to
$\xcf$ and $\xdf$. In the middle panel of \Fref{epsi} we plot the
foreground contamination parameter $\epsilon_\l$ in unit of
$\sigma_\l^\cmb$ against the cross correlation coefficients $\xdf$.
The thick line is for zero correlation between CMB and the
foregrounds:  $\xcf=0$, which is the case for high $\l$ with more
$m$ modes in the summation of \Eref{xcmbf}. For low $\l$, however,
the $\xcf$ plays a major role. In the bottom panel of \Fref{epsi} we
use the mean value of the $\xdfi$ from DILC and \wmap foreground
maps in \Fref{x} as input, and run $10^6$ Monte Carlo simulations for the $\xcf$. The foreground contamination is illustrated by the ratio $\epsilon_\l/\sigma_\l^\cmb$ at the values of 1, 2 and
3 $\sigma$ thresholds (yellow, light blue and dark blue areas, respectively). Note that the foreground contamination shown
in this way is independent of the underlying CMB power spectrum, and
we assume the foreground residuals in the DILC map has the same
morphology as the foregrounds themselves, so we are estimating an 
upper limit on the contamination ratio.

\section{Conclusion and Discussion}
In this paper we have proceeded from the basic assumption that the
CMB should have no or little correlation (within cosmic variance
limit) with the foregrounds, to examine foreground contamination in
the \wmap DILC map. We use the newly-developed representation 1DFR
of the spherical harmonic coefficients for cross and peak
correlation. The 1DFR has an advantage in that the $m=0$ modes do
not alter the morphology, and when using the standard definition of
cross correlation of the 1D curves, they are intrinsically excluded
in the calculation.

We have tested the DILC map for $2 \le \l \le 10$, although any
diagnostics of correlation can be applied to all $\l$ modes for both
the \wmap and the upcoming \planck data. We find that eight out of
nine modes are negatively-correlated, for which the probability is
only 0.0176 if they are really independent. We also find the local
extrema of the 1DFR curves are correlated, particularly for $\l=2$
and 6 modes, with significance level less than $0.04$. Both analyses
have indicated that the DILC quadrupole is contaminated with
foregrounds (with cross correlation coefficient $X< -0.8$ and extremum statistics at significance level $\alpha<0.04$). We then used the cross correlation coefficients to explore the level of the foreground
contamination.

The axiom that the CMB has no correlation with the foregrounds provides
the most fundamental criterion that should be applied as a foreground
cleaning check before any Gaussianity tests are performed and
cosmologists get carried away by suggestions of new physics. We
suggest that for the upcoming \planck mission, one should not ignore
the kind of information we have presented here, before any
scientific conclusion is reached.

One final remark for our results is that the DILC map is derived
from the concept of minimum variance optimization under the {\it a priori}
assumption that CMB and the foregrounds are statistically independent. That we
have only one universe, hence there is inevitably accidental correlation shall
introduce some error in the estimation of the weighting coefficients, and
consequently introduce residuals in the resultant map. Even if the correlation
between the CMB and the foregrounds happens to be zero, the variance of the
foreground residuals is minimum, but not necessarily zero. Although we can
extract the angular power spectrum as close as possible to that of the CMB
using such methods, the signal obtained in general still contains residuals.

\ack PC would like to thank the Niels Bohr International Academy for
their support during the Summer Institute when this paper was
written. We acknowledge the use of the NASA Legacy Archive for
extracting the \wmap data. We also acknowledge the use of \healpix
\footnote{\tt http://www.eso.org/science/healpix/} package
\cite{healpix} to produce $\alm$ from the \wmap data and the use of
\glesp \footnote{\tt http://www.glesp.nbi.dk} package \cite{glesp}.

\newcommand{\combib}[3]{\bibitem{#3}}       

%
%
\newcommand{\autetal}[2]{{#1\ #2 \etal}}    
\newcommand{\aut}[2]{{#1\ #2}}              
\newcommand{\saut}[2]{{#1\ #2},}             
\newcommand{\laut}[2]{{and #1\ #2},}         

%
%
\newcommand{\refsd}[6]{{\it #1}, #5 {\it #2} {\bf #3} #4 [#6]} 
\newcommand{\unrefs}[6]{{\it #1}, #5 {\it #2} #3 [#6]}         

%
%

\newcommand{\book}[6]{#5, #1, #2}                     

%
\newcommand{\proceeding}[6]{#5 {\it Proc. of #4} (#2) [#6]}   


\def\apj{\it Astrophys. J}
\def\apjl{\it Astrophys. J. Lett.}
\def\apjs{\it Astrophys. J. Supp.}
\def\ijmpd{\it Int. J. Mod. Phys. D}
\def\mn{\it Mon. Not. R. Astron. Soc.}
\def\nature{\it Nature}
\def\aa{\it Astro. \& Astrophys.}
\def\prl{\it Phys.\ Rev.\ Lett.}
\def\prd{\it Phys.\ Rev.\ D}
\def\pr{\it Phys.\ Rep.}

\def\cup{Cambridge University Press, Cambridge, UK}
\def\princetonpress{Princeton University Press}
\def\worldpress{World Scientific, Singapore}
\def\oxfordpress{Oxford University Press}

\end{document}